# Comment on "Chiral Phase Transition in Charge Ordered 1$T$-TiSe$_2$"

# and

# Supplementary Material on "First-order Forbidden X-ray Diffraction"

Meng-Kai Lin[1,2], Joseph A. Hlevyack[1,2], Peng Chen[1,2,3], Ro-Ya Liu[1,2,3,4], and T.-C. Chiang[1,2]

[1]*Department of Physics, University of Illinois at Urbana-Champaign, Urbana, Illinois 61801, USA*

[2]*Frederick Seitz Materials Research Laboratory, University of Illinois at Urbana-Champaign, Urbana, Illinois 61801, USA*

[3]*Advanced Light Source, Lawrence Berkeley National Laboratory, Berkeley, California 94720, USA*

[4]*Institute of Physics, Academia Sinica, Taipei 10617, Taiwan*

A prior report of the emergence of chirality for the (2x2x2) charge density wave (CDW) in TiSe$_2$ has attracted much interest; the drastic symmetry breaking is highly unusual with few precedents [1]. In that study, key evidence was provided by x-ray diffraction measurements of two superlattice reflections, (1.5 1.5 0.5) and (2.5 1 0). The (2.5 1 0) reflection appeared to show an anomalously large intensity and a transition onset at ~7 K below that of the (1.5 1.5 0.5) reflection. These observations, aided by modeling, were cited as evidence for a separate chiral transition. In this Comment, we show that the prior conclusions based on x-ray diffraction are erroneous. There is just one transition, and it is achiral.

Figure 1 shows x-ray diffraction intensity data reproduced from Ref. [1] for the two CDW reflections, (1.5 1.5 0.5) and (2.5 1 0). Also shown are calculations based on the x-ray structure factors derived from known quantities for the CDW without any assumed chirality [2-4]. There are no adjustable parameters in the calculation except for (1) an overall intensity scale factor, which is chosen to match the calculation to both sets of data simultaneously, and (2) an adjustment of the CDW transition temperature from 205 to 190 K to account for the somewhat different sample stoichiometry. The calculations agree well with the experiment.

The two curves in Fig. 1 have very different shapes near the onset. The (1.5 1.5 0.5) intensity can be well described by a linear approximation, characteristic of a second-order phase transition. The much weaker (2.5 1 0) peak appears to exhibit a delayed onset and thus a lower transition temperature; this was attributed to a chiral transition at ~7 K below the CDW transition in Ref. [1]. However, this CDW peak, with a zero momentum transfer in the layer normal direction, has a zero structure factor to first order because contributions from the two neighboring layers in a (2x2x2) unit cell with opposite CDW distortions cancel [5]. As a result, the intensity rise should be quadratic below the onset. The onset temperature is actually the same, but the quadratic dependence of the intensity may give the impression of a delayed onset. According to Ref. [1], a chiral order should give rise to a very large enhancement of the (2.5 1 0) structure factor by a factor of 1.86 and 64 for the Ti and Se contributions, respectively. The weighted average of the enhancement factor is thus 43, a huge effect. Our analysis shows that the experimentally measured intensity relative to that of the (1.5 1.5 0.5) peak agrees with an achiral CDW with no evidence for any enhancement.

Reference [1] also contained data of specific heat and resistivity measurements. Analysis therein suggested a second transition, but the features are weak, broad, and not very well-defined. Other prior studies on this topic include scanning tunneling microscopy and optical measurements with conflicting conclusions [6, 7]. The present Comment focuses on the x-ray data only. X-ray diffraction, being highly sensitive to atomic movements, is the most direct tool to discern details of such transitions at the atomic level. We conclude that the CDW in $TiSe_2$ involves a single achiral transition to a high level of precision.

This work is supported by the U.S. Department of Energy (DOE), Office of Science (OS), Office of Basic Energy Sciences, Division of Materials Science and Engineering, under Grant No. DE-FG02-07ER46383.

FIG. 1 Experimental and calculated x-ray intensities for the (1.5 1.5 0.5) and (2.5 1 0) reflections as a function of temperature. The intensity of the (2.5 1 0) reflection is much weaker and is amplified by a factor of 50 for both the data and the calculation.

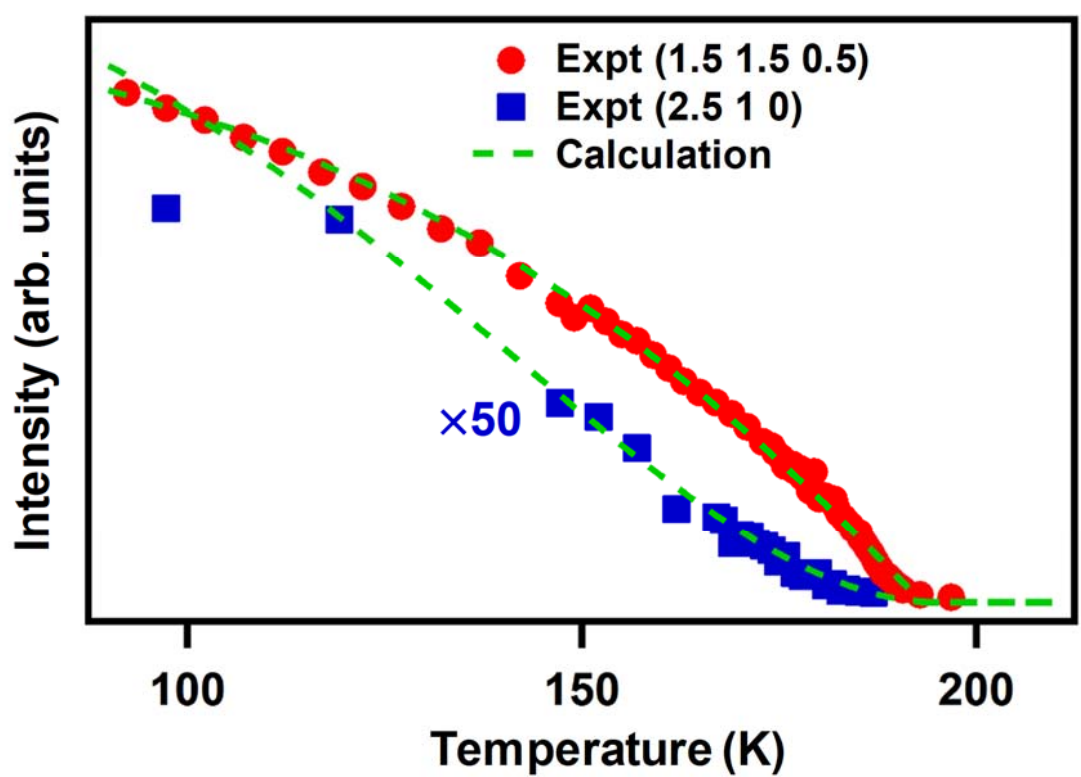

## Supplementary Material

## First-order Forbidden X-ray Diffraction

The structure factor for TiSe$_2$ in the (2x2x2) CDW phase at scattering vector $\mathbf{q} = (H\ K\ L)$ in reciprocal lattice units is given by

$$S(\mathbf{q}) = \sum_j f_j(\mathbf{q}) \exp(-i\mathbf{q} \cdot (\mathbf{R}_j + \mathbf{u}_j)) \tag{1}$$

where $j$ is an index for the atoms within a (2x2x2) unit cell, $f_j$ the atomic form factor, $\mathbf{R}_j$ the undistorted atomic position, and $\mathbf{u}_j$ the atomic displacement in the CDW phase. Expanding the exponential function as a power series of $\mathbf{u}_j$, one obtains

$$S = S_0 + S_1 + S_2 + \ldots = \sum_j f_j \exp(-i\mathbf{q} \cdot \mathbf{R}_j) - i\mathbf{q} \cdot \sum_j \mathbf{u}_j f_j \exp(-i\mathbf{q} \cdot \mathbf{R}_j) + O(u^2) + \ldots \tag{2}$$

$S_0$, structure factor for the undistorted (2x2x2) structure, is zero for half-order reflections such as (1.5 1.5 0.5) and (2.5 1 0). The first-order term in Eq. (2) contains the factor

$$\sum_j \mathbf{u}_j f_j \exp(-i\mathbf{q} \cdot \mathbf{R}_j) = \sum_j^T \mathbf{u}_j f_j \exp(-i\mathbf{q} \cdot \mathbf{R}_j) + \sum_j^B \mathbf{u}_j f_j \exp(-i\mathbf{q} \cdot \mathbf{R}_j) \tag{3}$$

where it is split into two parts. The first (second) part involves a summation over the top (bottom) (2x2) unit in the (2x2x2) unit cell.

For the (2.5 1 0) reflection, $q_z = 0$, and the phase factor $\exp(-i\mathbf{q} \cdot \mathbf{R}_j)$ in Eq. (3) is independent of the $z$-coordinates of the atoms. The two parts in Eq. (3) cancel each other because each of the atomic displacements in the top (2x2) unit equals the corresponding one in the bottom (2x2) unit but in the opposite direction. This anti-phase displacement pattern is responsible for the CDW period doubling along the layer normal. As a result, the (2.5 1 0) reflection is forbidden to first order. The first nonvanishing term for $S$ is second order in the atomic displacements.

$$S(2.5\ 1\ 0) = O(u^2) \tag{4}$$

This cancellation does not apply for the (1.5 1.5 0.5) reflection, and the structure factor is dominated by the first-order terms.

$$S(1.5\ 1.5\ 0.5) = O(u) \tag{5}$$

For the second-order CDW transition in TiSe$_2$, the allowed reflections such as (1.5 1.5 0.5) should show a linear rise in x-ray intensity below but near the transition temperature $T_C$ (see Ref. [2] for details):

$$I(1.5\ 1.5\ 0.5) \propto \left(1 - \frac{T}{T_C}\right) \Theta(T_C - T) \tag{6}$$

The first-order forbidden reflections such as (2.5 1 0) should show a quadratic rise instead:

$$I(2.5\ 1\ 0) \propto \left(1 - \frac{T}{T_C}\right)^2 \Theta(T_C - T) \tag{7}$$

It is straightforward to show that all half-order reflections with $L$ being an integer (not just zero) are forbidden to first order. The calculated intensities shown in Fig. 1 were obtained numerically using Eq. (1), the known atomic form factors, and the temperature dependence of the order parameter for a second-order phase transition [2].